\documentclass[final]{svjour3}
\usepackage{graphicx}
\usepackage{rotating}
\usepackage{amssymb}
\usepackage{mathptmx}
\usepackage{multirow}
\usepackage{hyperref}
\usepackage{threeparttable}
\makeatletter
\journalname{Journal of Low Temperature Physics}
\usepackage{cite}

\begin{document}

\newcommand{\hdblarrow}{H\makebox[0.9ex][l]{$\downdownarrows$}-}
\title{The CLASS 150/220 GHz Polarimeter Array: Design, Assembly, and Characterization}

\author{S. Dahal \and M. Amiri \and J. W. Appel \and C. L. Bennett \and L. Corbett \and R. Datta \and K. Denis \and T. Essinger-Hileman \and M. Halpern \and K. Helson \and G. Hilton \and J. Hubmayr \and B. Keller \and T. Marriage \and C. Nunez \and M. Petroff \and C. Reintsema \and K. Rostem \and K. U-Yen \and E. Wollack  }

\institute{S. Dahal \and J. W. Appel \and C. L. Bennett \and L. Corbett \and R. Datta \and B. Keller \and T. Marriage \and C. Nunez \and M. Petroff 
            \at Department of Physics and Astronomy, Johns Hopkins University, Baltimore, MD 21218, USA\\ \email{sumit.dahal@jhu.edu}
            \and M. Amiri \and M. Halpern
            \at Department of Physics and Astronomy, University of British Columbia, Vancouver, BC V6T 1Z4, Canada
            \and K. Denis \and T. Essinger-Hileman \and K. Helson \and K. Rostem \and K. U-Yen \and E. Wollack  
            \at NASA Goddard Space Flight Center, Greenbelt, MD 20771, USA
            \and G. Hilton \and J. Hubmayr \and C. Reintsema
            \at National Institute of Standards and Technology, Boulder, CO 80305, USA}

\maketitle
\begin{abstract}
We report on the development of a polarization-sensitive dichroic (150/220 GHz) detector array for the Cosmology Large Angular Scale Surveyor (CLASS) delivered to the telescope site in June 2019. In concert with existing 40 and 90 GHz telescopes, the 150/220~GHz telescope will make observations of the cosmic microwave background over large angular scales aimed at measuring the primordial B-mode signal, the optical depth to reionization, and other fundamental physics and cosmology. The 150/220 GHz focal plane array consists of three detector modules with 1020 transition edge sensor (TES) bolometers in total. Each dual-polarization pixel on the focal plane contains four bolometers to measure the two linear polarization states at 150 and 220 GHz. Light is coupled through a planar orthomode transducer (OMT) fed by a smooth-walled feedhorn array made from an aluminum-silicon alloy (CE7). In this work, we discuss the design, assembly, and in-lab characterization of the 150/220 GHz detector array.  The detectors are photon-noise limited, and we estimate the total array noise-equivalent power (NEP) to be 2.5 and 4 aW$\sqrt{\mathrm{s}}$ for 150 and 220 GHz arrays, respectively.

\keywords{CMB polarization, CLASS, dichroic, TES}

\end{abstract}

\section{Introduction}
Precise measurement of the cosmic microwave background (CMB) polarization at large angular scales enables us to characterize cosmic inflation, pinpoint the epoch of reionization, constrain the mass of neutrinos, and probe new physics beyond the standard model including early dark energy models intended to resolve the Hubble constant problem. However, characterizing the faint polarized CMB signal is challenging due to large polarized foreground contamination at millimeter wavelengths from Galactic synchrotron and dust emission. It is therefore essential to pursue multi-band observations using large arrays of efficient background-limited detectors with excellent control over instrumental systematics.

\begin{figure}[h]
   \begin{center}
   \includegraphics[scale=0.065]{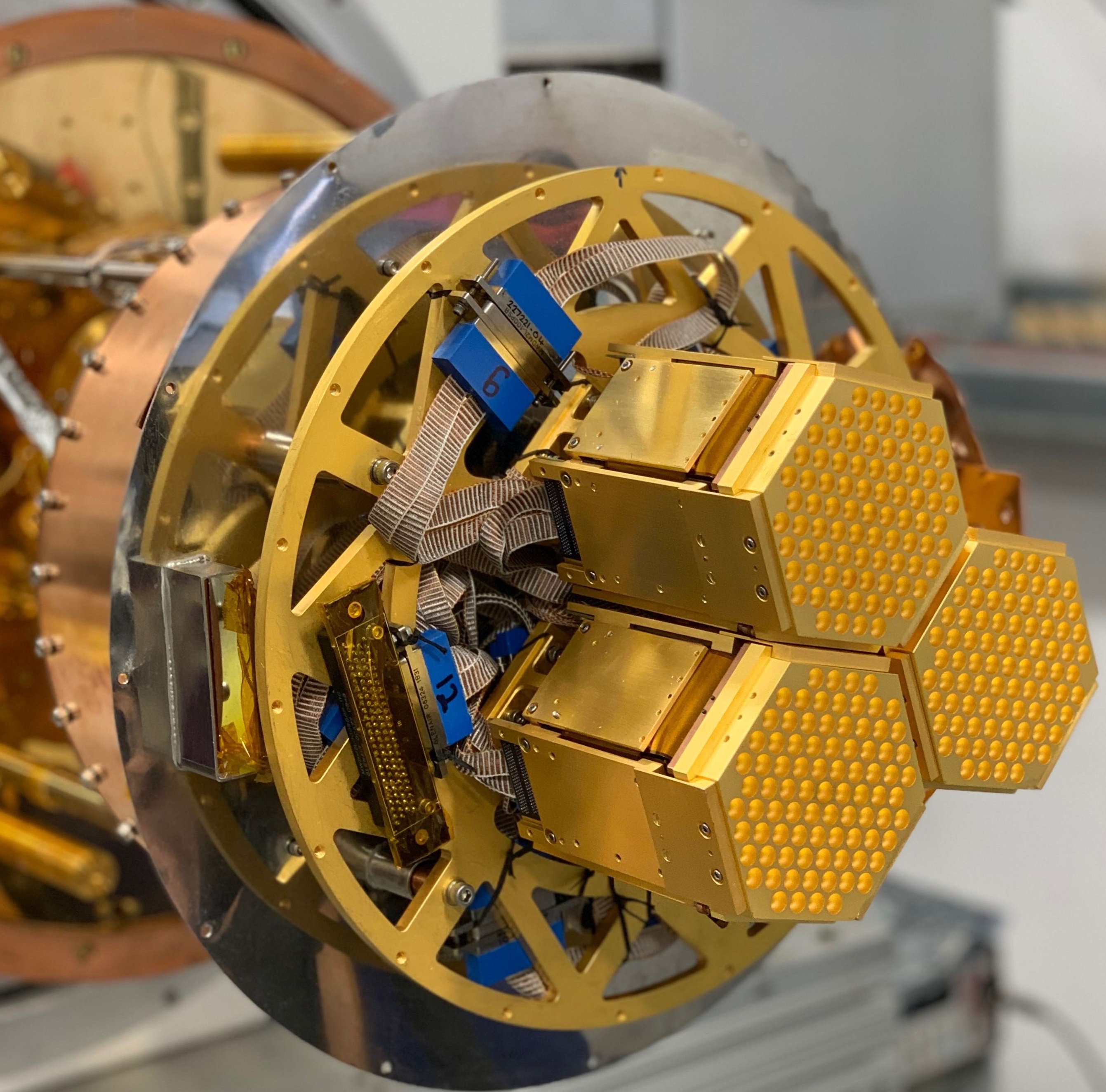}
   \end{center}
   \caption{The HF detector array at the CLASS telescope site in Chile during testing. The array consists of three identical hexagonal modules mounted onto the mixing chamber plate of a pulse-tube cooled dilution refrigerator using a Au-coated copper web interface seen here. There are 1020 polarization-sensitive TES bolometers on the focal plane split equally between 150 and 220 GHz. (Color figure online.)} 
   \label{fig:fp} 
   \end{figure}

The Cosmology Large Angular Scale Surveyor (CLASS) maps the microwave sky from the Atacama Desert in Chile at four frequency bands from 40 to 220 GHz, optimized for maximum signal-to-noise based on the atmospheric emission.  The addition of a dichroic (150/220 GHz) detector array to the existing 40 \cite{john19} and 90 GHz \cite{sumit} telescopes provides additional sensitivity to CLASS's CMB observations and helps to characterize the dust foreground. The 150/220 GHz high frequency (HF) instrument was delivered to and tested at the CLASS site in June 2019. The HF focal plane array consists of three detector modules, seen in Figure \ref{fig:fp}, with 255 dichroic dual-polarization pixels in total. Each pixel consists of four transition edge sensor (TES) bolometers that couple to light through a planar orthomode transducer (OMT) \cite{kevin,karwan} fed by a smooth-walled feedhorn array made from an aluminum-silicon Controlled Expansion 7 (CE7) alloy.\footnote{www.smt.sandvik.com/osprey} Here we describe the design and assembly, and present lab characterization of the HF detector array.

\section{Detector Design}
\label{sec:design}
\begin{figure}
   \begin{flushleft}
   \includegraphics[trim = 0 2.2cm 2.1cm 1.2cm, clip=true, scale=0.38]{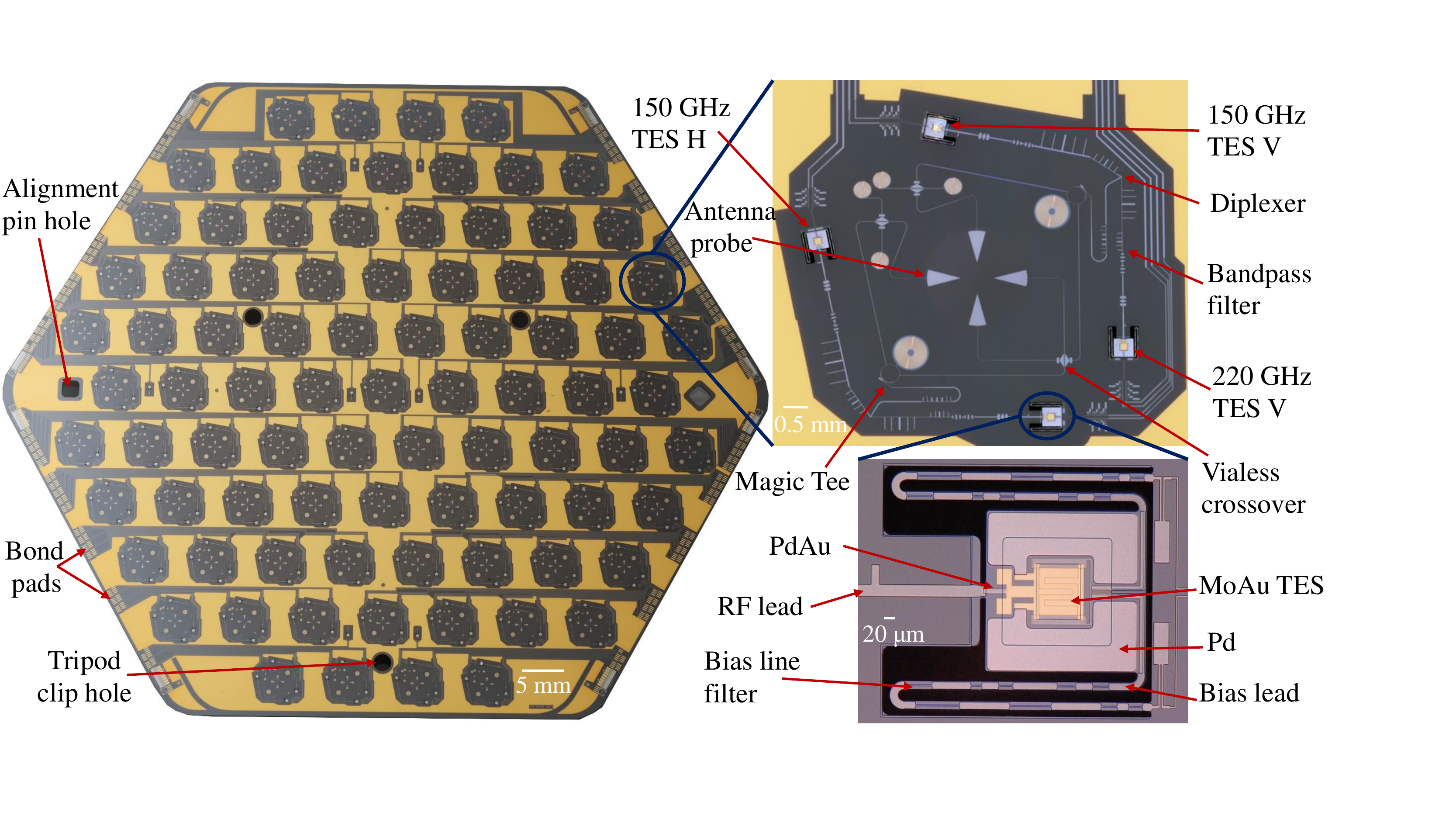}
   \end{flushleft}
   \caption{\textbf{Left:} CLASS HF detector wafer with 85 dichroic dual-polarization pixels fabricated on a monocrystalline silicon layer. Detector readout signals are routed to the bond pads located near four edges of the wafer. \textbf{Right:} Zoomed-in image of a single detector pixel (top) and a TES island (bottom). The optical signal on the microstrip transmission lines coming from the OMTs is separated into two bands by a diplexer plus on-chip filters and terminated on the TES bolometers. For a single frequency band, the detector architecture is similar to the CLASS 90~GHz design described in \cite{kevin,karwan}. (Color figure online.)} 
   \label{fig:wafer} 

   \end{figure}

The CLASS HF detectors are fabricated on 100 mm silicon wafers each consisting of 85 dichroic polarization-sensitive pixels as shown in Figure \ref{fig:wafer}. The beam is defined by a smooth-walled CE7 feedhorn, which couples light onto an OMT. The OMT separates two orthogonal states of linear polarization and couples them to microstrip transmission lines. The signals from opposite OMT probes are combined onto a single microstrip line using the difference output of a magic tee \cite{kpop}. A diplexer then splits each signal path into two, followed by on-chip filters that define the frequency band. The signal continues on the microstrip line to the TES membrane through a stubby beam and is terminated at a PdAu resistor. The stubby beam precisely controls the thermal conductance of the TES island with ballistic-dominated phonon transport \cite{karwan14}. The long and meandered legs that support the TES bias leads have rough side walls and contribute very little to the thermal conductance. The Pd seen in the TES island in Figure \ref{fig:wafer} is used to define the detector time constant through the electronic heat capacity of Pd.

Similar to the 90 GHz architecture described in \cite{sumit,kevin,karwan}, the HF wafer package consists of three separate wafers hybridized together: a photonic choke wafer, a detector wafer, and a backshort assembly. The photonic choke \cite{wollack} acts as a waveguide interface between the feedhorn array and the detectors. The backshort assembly \cite{crowe}, among its many functions \cite{karwan}, forms a quarter-wavelength short for the OMT antenna probes. The middle detector wafer contains the detectors fabricated on single-crystal silicon that has excellent microwave and thermal properties \cite{karwan}. The hybridized wafer package is mounted to the CE7 feedhorn array as a single assembly.

\section{Module Design and Assembly}
\begin{figure}
   \begin{flushleft}
   \includegraphics[trim = 0cm 5.2cm 2cm 7cm, clip=true, scale=0.35]{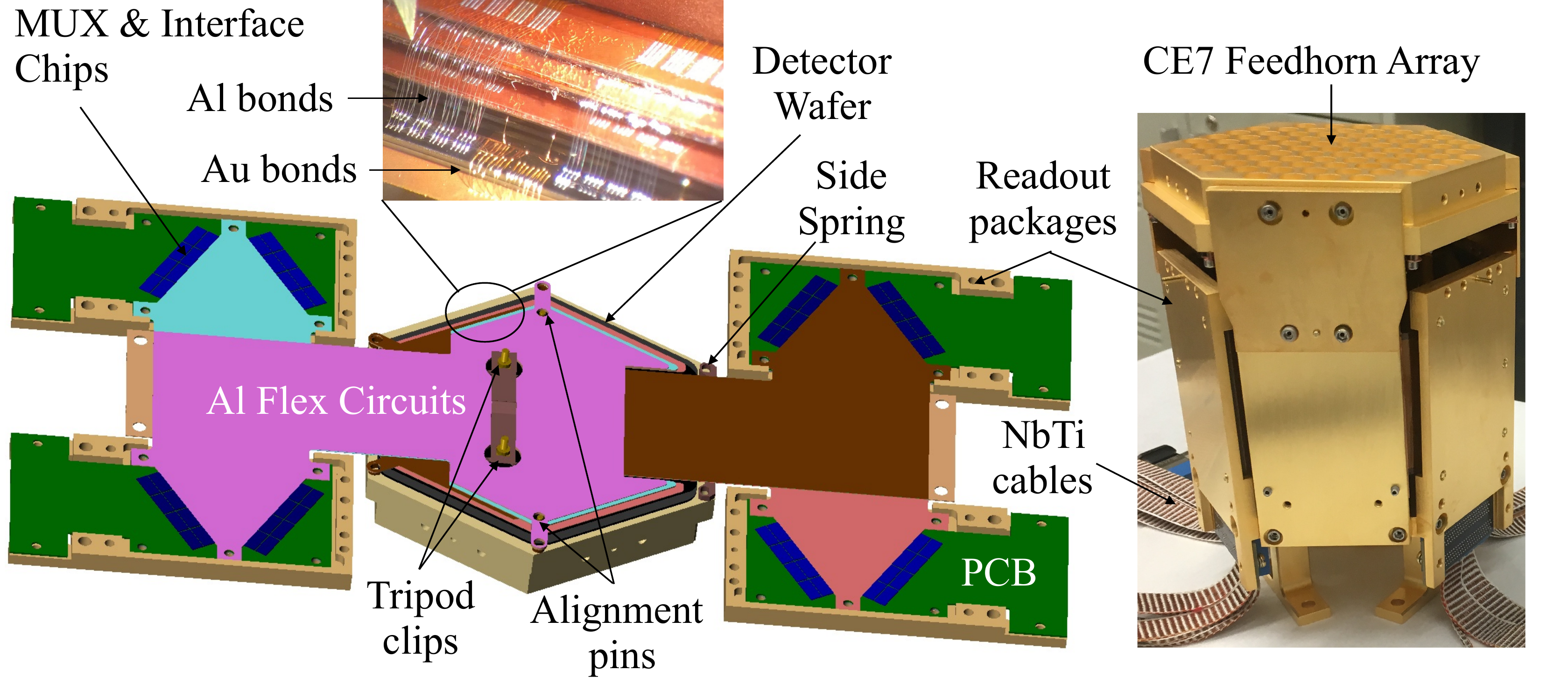}
   \end{flushleft}
   \caption{\textbf{Left:} Model of unfolded HF module during assembly. The detector wafer (black) is mounted on top of Au-plated CE7 feedhorn array using two Be-Cu tripod clips. Four layers of Al flex circuits with decreasing circumradius (starting from bottom: coral, brown, light blue, and pink) are stacked on top of the wafer and connected to separate readout packages. These packages contain MUX and interface chips (blue) mounted onto a PCB (green) sandwiched between two Nb sheets (not shown). The inset shows intricate layers of Al bonds from the wafer to different flex circuit layers (the topmost layer is not visible here). Au bonds heat sink the detector wafer to the feedhorn array. \textbf{Right:} An assembled HF module. After assembly, all four readout packages are folded up and bolted to the CE7. Support structures are bolted to the bottom through a backplate. (Color figure online.)} 
   \label{fig:module}
   \end{figure}

As shown in Figure \ref{fig:fp}, the HF focal plane consists of three identical modules mounted on a mixing chamber plate maintained at a stable bath temperature of $\sim$~80~mK by a pulse-tube cooled dilution refrigerator.\footnote{BlueFors Cryogenics, www.bluefors.com} Each module consists of 85 smooth-walled feedhorns made from a Au-plated CE7 alloy, which is machinable and has lower differential contraction to silicon detector wafers than copper or aluminum \cite{aamir}. Figure \ref{fig:module} shows the tripod clips, alignment pins, and the side spring used to mount and align the detector wafer onto the feedhorn array. During assembly, four layers of flexible aluminum circuits are stacked on top of each wafer package and wire bonded to the detector bond pads on four sides of the wafer. The other end of the circuits are mounted onto four separate readout packages seen in Figure \ref{fig:module}.

Each readout package consists of eight time-division multiplexer (MUX) chips \cite{nist} and eight interface chips mounted on a printed circuit board (PCB), sandwiched between two Nb sheets for magnetic shielding. The interface chips have 200 $\mathrm{\mu \Omega}$ shunt resistors and 310~nH Nyquist inductors. The inductance value was chosen to keep the readout noise aliased from higher frequencies below one percent of the noise level within the audio bandwidth of the TES. We also performed tests with and without this inductor to ensure that its addition does not affect the detector stability. Four MUX chips on each side of a readout package are combined into one readout column for multiplexing 44 rows i.e. a multiplexing factor, or number of detectors per readout channel, of 44:1. The 150 and 220 GHz detectors are mapped to separate readout packages, each with two columns, so that they can be biased separately, if needed.  

Detector bias signals and MUX addressing, feedback, and output signals are routed via twisted pairs of NbTi superconducting cables (seen in Figure \ref{fig:module}), soldered directly into the PCB. In particular, this cabling connects the MUX output to a set of SQUID series array (SSA) amplifiers. The 4K SSA and the warm electronics are identical to the CLASS 90~GHz components as described in \cite{sumit}. Figure \ref{fig:module} shows the layout of a module during and after assembly. After all the components are assembled and wire bonded, the readout packages are folded up and bolted to the feedhorn array on one side and attached to a backplate on the other. The backplate is finally bolted to the mixing chamber plate with the help of a web structure and posts as seen in Figure \ref{fig:fp}. 

\section{Detector Characterization}
\label{sec:charac}
\subsection{Electro-thermal Parameters}
\label{sec:parameters}
We characterize electro-thermal properties of the detectors by capping off all the cold stages of the cryostat with metal plates so as to minimize optical loading. We measure the TES saturation power ($P_\mathrm{sat}$) at 80\% TES normal resistance (R$_{\mathrm{N}}$) for multiple bath temperatures ($T_\mathrm{bath}$) from 70 to 250 mK through I-V curves. At each $T_\mathrm{bath}$, we ramp up the voltage bias to drive all the detectors normal, then sweep the bias down through the superconducting transition while recording the current response of the detectors. We fit $P_\mathrm{sat}$ vs $T_\mathrm{bath}$ to the following power-law:
\begin{equation}
\label{eq:psat}
P_{\mathrm{sat}} = \kappa \ (T_\mathrm{c}^n - T_{\mathrm{bath}}^n)\ ,
\end{equation}
where $\kappa$ is a thermal conductance parameter, and $n$ is a power law index governing the power flow between the TES island and the bath. For CLASS detectors, this power flow is dominated by ballistic phonons, so we fix $n$ = 4 \cite{karwan14}. Table \ref{tab:param} shows the mean and standard deviation of $T_{\mathrm{c}}$ and $\kappa$ values obtained from the fit to Equation \ref{eq:psat} for three CLASS HF modules. The $P_\mathrm{sat}$ value is calculated at $T_\mathrm{bath}$ = 50 mK, and the yield is the fraction of detectors that fit Equation \ref{eq:psat} with median absolute deviation (MAD) $<$ 1 pW. Detectors with MAD~$>$~1~pW are not included in the mean and standard deviation calculations for Table \ref{tab:param} and are excluded from further analysis in this paper.

\begin{table}[ht]
\caption{Mean and standard deviations of detector parameters and yield for the three CLASS HF modules} 
\label{tab:param}
\begin{center}       
\begin{tabular}{c|c|c|c|c|c|c} 
\textbf{Module} & \textbf{$\nu$ (GHz)} & \textbf{$T_{\mathrm{c}}$ (mK)} & \textbf{$\kappa$ (nW/K$^4$)} & \textbf{$P_\mathrm{sat}$ (pW)} & \textbf{NEP$_G$ (aW$\sqrt{\mathrm{s}}$)} & \textbf{Yield (\%)} \\
\hline
\hline
\multirow{2}{*}{1}& 150 & 214 $\pm$ 10 & 18 $\pm$ 1 & 39 $\pm$ 6 & 21 $\pm$ 2 & 81 \\
& 220 & 213 $\pm$ 10 & 22 $\pm$ 1 & 44 $\pm$ 6 & 23 $\pm$ 2 & 70 \\
\hline
\multirow{2}{*}{2}& 150 & 199 $\pm$ 7 & 20 $\pm$ 2 & 31 $\pm$ 4 & 18 $\pm$ 2 & 76 \\
& 220 & 200 $\pm$ 5 & 22 $\pm$ 2 & 36 $\pm$ 3 & 20 $\pm$ 1 & 41 \\
\hline
\multirow{2}{*}{3}& 150 & 204 $\pm$ 13 & 20 $\pm$ 1 & 34 $\pm$ 7 & 20 $\pm$ 3 & 82 \\
& 220 & 210 $\pm$ 10 & 23 $\pm$ 1 & 44 $\pm$ 6 & 23 $\pm$ 2 & 61 \\
\hline
\hline
\multirow{2}{*}{\textbf{Total}}& 150 & 206 $\pm$ 12 & 19 $\pm$ 2 & 35 $\pm$ 7 & 20 $\pm$ 3 & 80 \\
& 220 & 209 $\pm$ 10 & 22 $\pm$ 2 & 42 $\pm$ 7 & 22 $\pm$ 2 & 57 \\
\end{tabular}
\end{center}
\end{table}

The differences in $P_\mathrm{sat}$ values obtained with increasing vs decreasing bath temperatures were $\lesssim$ 0.2 pW; therefore, measurement errors were ignored in the standard deviation values of Table \ref{tab:param}. As seen in Table \ref{tab:param}, the spreads of $T_{\mathrm{c}}$ (3 -- 6\%) and $\kappa$ (4 -- 10\%) parameters across all three modules for both frequencies are small. This results from uniform and controlled fabrication processes, and ballistic thermal transport to the bath in all the TESs. The spread in $P_\mathrm{sat}$ (8 -- 20\%) is mostly explained by the spread in $T_{\mathrm{c}}$. The uniformity of the detector parameters across the array and the observed in-lab detector stability over a wide range of bias voltages will  allow the array to be optimally biased during  sky observations. At the time of writing, the warm readout Multichannel Electronics (MCE) \cite{mce} firmware supports multiplexing over a maximum of 41 rows. Through a firmware upgrade, we expect to get all 44 rows working, which will increase the 150 GHz array yield to $\sim$ 86\% and 220~GHz to $\sim$~61\%. Most of the remaining detectors are not operational due to complications in the bonding geometry. In particular, the lower yield in 220 GHz versus 150 GHz is due to wire bonding difficulty when trying to bond to the upper layers of the Al flex circuit stack as shown in the inset of Figure \ref{fig:module}.

\subsection{Bandpass}
\label{sec:bandpass}
\begin{figure}
\begin{center}
\includegraphics[scale=0.24]{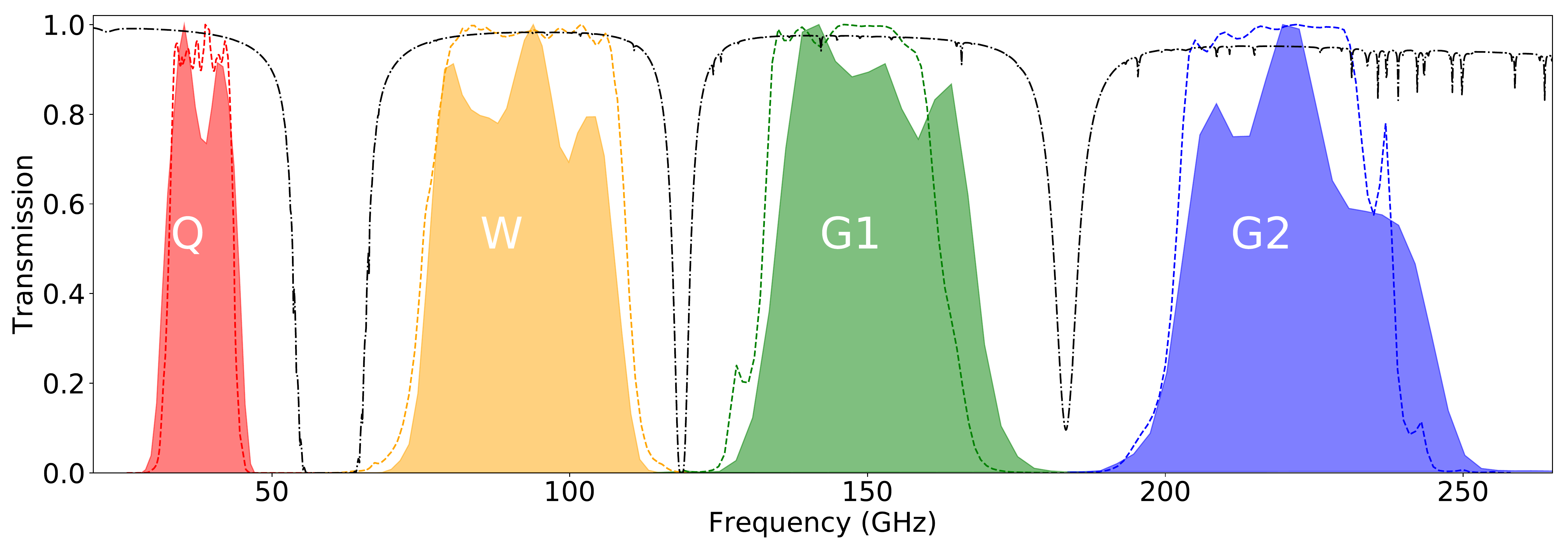}
\end{center}
\caption{Measured bandpasses (filled) of CLASS detector arrays compared to simulation (dashed) and atmospheric transmission model (dash-dot) \cite{atmosphere} at the CLASS site with PWV of 1 mm. The bandpasses were measured in lab with a polarizing Martin-Puplett Fourier-transform spectrometer and have been corrected for the feedhorn's frequency-dependent gain and the transmission through cryostat filters. (Color figure online.)}
\label{fig:bandpass}
\end{figure}

We use a polarizing Martin-Puplett Fourier Transform Spectrometer (FTS) to measure the bandpasses of the HF detectors in lab. Refer to \cite{fts} for detailed description of the FTS used and \cite{sumit} for the filters used in the lab-cryostat setup. Figure \ref{fig:bandpass} shows the measured CLASS detector bandpasses compared to simulation and the atmospheric transmission model at the CLASS site in the Atacama Desert with precipitable water vapor (PWV) of 1 mm. The raw measured bandpass values have been corrected for the feedhorn's frequency-dependent gain and the transmission through the cryostat filters used in the lab setup.

 The measured CLASS bandpasses in Figure \ref{fig:bandpass} have been averaged over all working detectors in the band. The CLASS Q and W bandpasses are added for completeness and are described further in \cite{john19} and \cite{sumit}, respectively. The bandpasses safely avoid strong atmospheric emission lines, as designed. We also do not see any evidence of high frequency out-of-band leakage. The simulated bandpasses for the two HF bands are 132 -- 162 GHz and 202 -- 238 GHz as defined by their half-power points; whereas the respective measured bands are 134 -- 168~ GHz and 203 -- 241 GHz. So, compared to simulation, the measured bandpasses for 150 and 220~GHz detectors are wider by 4 and 2 GHz, respectively, and both bands are shifted by a few GHz toward higher frequencies. We are currently investigating the source of these apparent shifts in the bands, which may be due to FTS systematics. We also plan to make FTS measurements through the receiver optics in the field.

\subsection{Noise and Sensitivity}
\label{sec:noise}
\begin{figure}
\begin{flushleft}
\includegraphics[scale=0.287]{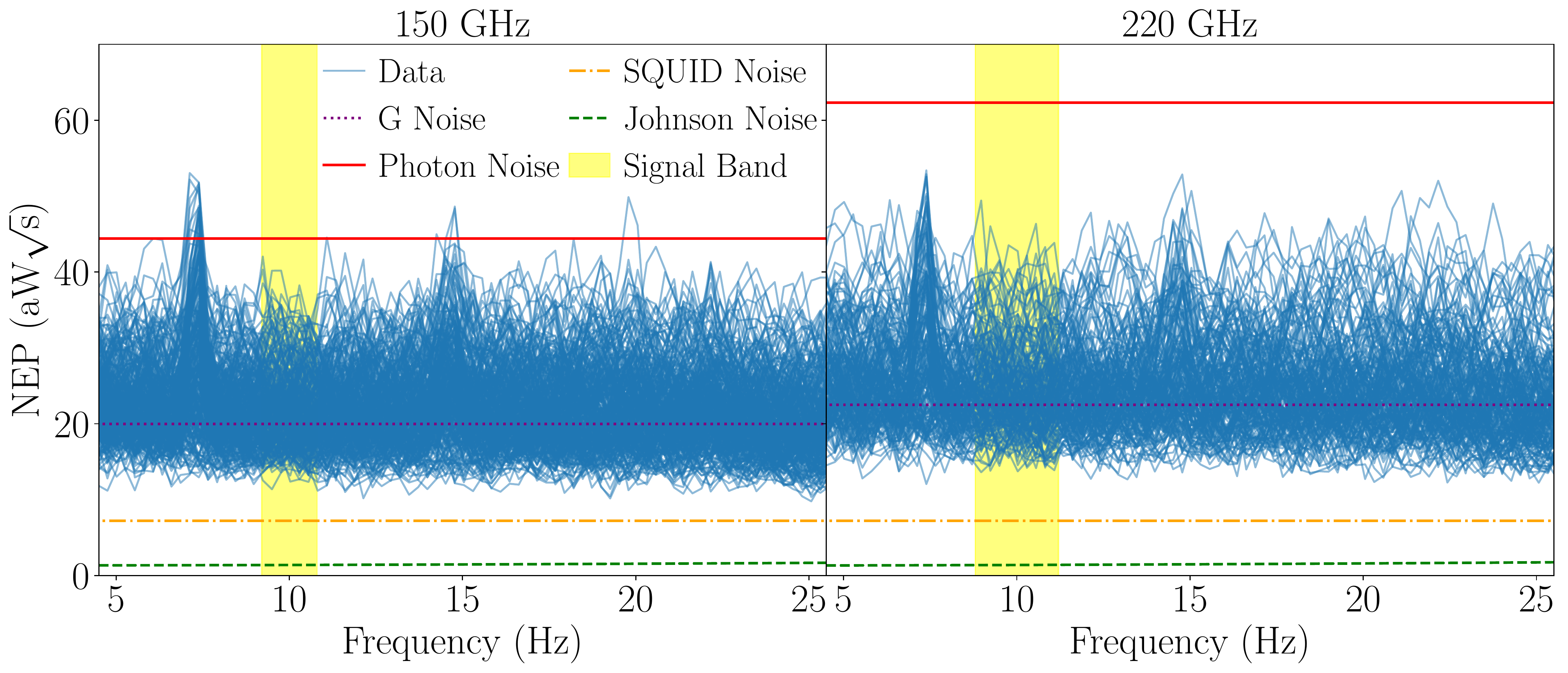}
\end{flushleft}
\caption{Noise spectra of CLASS HF detectors operated in the dark. The horizontal lines show the NEP$_\mathrm{dark}$ components and estimated photon noise. The vertical yellow patch shows the CLASS audio signal band centered at the VPM modulation frequency of 10 Hz. The measured average NEP of 22 aW$\sqrt{\mathrm{s}}$ for 150 GHz and 25~aW$\sqrt{\mathrm{s}}$ for 220 GHz matches well with the expected G noise (from Table \ref{tab:param}) as the SQUID noise and the Johnson noise are negligible when added in quadrature. Given the noise spectra and estimated photon noise, all the working HF detectors are photon-noise limited. (Color figure online.)}
\label{fig:nep}
\end{figure}

We took noise spectra of the HF detectors ``in the dark'' (capping off all the cold stages of the cryostat with metal plates) and measured the detector dark noise-equivalent power (NEP$_\mathrm{dark}$) shown in Figure \ref{fig:nep}. In the CLASS audio signal band centered at the variable-delay polarization modulator (VPM) \cite{katie} frequency of 10 Hz, the mean and standard deviation of NEP$_\mathrm{dark}$ is 22 $\pm$ 4 aW$\sqrt{\mathrm{s}}$ and 25 $\pm$ 4 aW$\sqrt{\mathrm{s}}$ for 150 and 220 GHz detectors, respectively. TES phonon noise (NEP$_\mathrm{G}$), detector Johnson noise, and SQUID readout noise contribute to the measured NEP$_\mathrm{dark}$. We measured the current noise of ``dark SQUIDs'' (SQUIDs that are not connected to detectors) to be $\sim$ 35 pA$\sqrt{\mathrm{s}}$. Using an average detector responsivity of $\sim \ 5 \times 10^6 \  \mathrm{V^{-1}}$, the SQUID readout noise is 7 aW$\sqrt{\mathrm{s}}$. The detector Johnson noise is highly suppressed through electro-thermal feedback at lower frequencies, and we estimate it to be $\sim$ 1.5 aW$\sqrt{\mathrm{s}}$. The expected NEP$_\mathrm{G}$ values calculated from the detector parameters are shown in Table \ref{tab:param}. NEP$_\mathrm{G}$ is the dominant noise source for NEP$_\mathrm{dark}$ as the readout and the Johnson noise contribute only a few percent when added in quadrature.             

As shown in Figure \ref{fig:nep}, the estimated photon NEP in the field for the 150 and 220 GHz detectors are 44.4 and 62.3 aW$\sqrt{\mathrm{s}}$, respectively \cite{tom}. Given the measured NEP$_\mathrm{dark}$ values, all the working detectors on the HF array are photon-noise limited. With the current array yield, the total array NEP is 2.5 aW$\sqrt{\mathrm{s}}$ for 150 GHz and 4 aW$\sqrt{\mathrm{s}}$ for 220 GHz. Assuming nominal bandpass and 50\% total optical efficiency, we estimate array NETs of 17 and 51~$\mu$K$\ \sqrt[]{\mathrm{s}}$ for 150 and 220 GHz, respectively. Since the HF VPM is optimized for 150 GHz, we estimate the modulation efficiency to
measure the Stokes parameter Q to be 70\% for 150~GHz and 50\% for 220 GHz. This leads to array noise-equivalent Q (NEQ) of 24 $\mu$K$\ \sqrt[]{\mathrm{s}}$ for 150 GHz and 101 $\mu$K$\ \sqrt[]{\mathrm{s}}$ for 220 GHz.

\section{Conclusion}
The CLASS HF detector array has been delivered to the telescope site and installed inside the cryostat receiver. This dichroic array sensitive to 150 and 220 GHz will provide additional sensitivity to CLASS's CMB observations and help characterize the dust foreground. The detectors within each HF module show uniform parameter distributions. FTS measurements performed in lab show that the detector bandpasses safely avoid strong atmospheric emission lines with no evidence for high frequency out-of-band leakage. The HF detectors are photon-noise limited with average NEP$_\mathrm{dark}$ of 22 aW$\sqrt{\mathrm{s}}$ for 150 GHz and 25~aW$\sqrt{\mathrm{s}}$ for 220~GHz. With current array yield and expectations for optical and VPM modulation efficiencies, we estimate the CLASS HF array NEQ of 24 $\mu$K$\ \sqrt[]{\mathrm{s}}$ for 150 GHz and 101 $\mu$K$\ \sqrt[]{\mathrm{s}}$ for 220 GHz.

\begin{acknowledgements}
We acknowledge the National Science Foundation Division of Astronomical Sciences for their support of CLASS under Grant Numbers 0959349, 1429236, 1636634, and 1654494. CLASS uses detector technology developed under several previous and ongoing NASA grants. Detector development work at JHU was funded by NASA grant number NNX14AB76A. We further acknowledge the very generous support of Jim and Heather Murren (JHU A\&S '88), Matthew Polk (JHU A\&S Physics BS '71), David Nicholson, and Michael Bloomberg (JHU Engineering  '64). CLASS is located in the Parque Astron\'{o}mico de Atacama in northern Chile under the auspices of the Comisi\'{o}n Nacional de Investigaci\'{o}n Cient\'{i}fica y Tecnol\'{o}gica de Chile (CONICYT).
\end{acknowledgements}


\begin{thebibliography}{12}
\bibitem{john19}
J. W. Appel et al., {\it ApJ} \textbf{876}, 126, (2019), DOI: 10.3847/1538-4357/ab1652

\bibitem{sumit}
S. Dahal et al., {\it Proc. SPIE} \textbf{10708}, 107081Y, (2018), DOI: 10.1117/12.2311812

\bibitem{kevin}
K. L. Denis et al., {\it AIP Conference Proc.} \textbf{1185}, 371, (2009), DOI: 10.1063/1.3292355

\bibitem{karwan}
K. Rostem et al., {\it Proc. SPIE} \textbf{9914}, 99140D, (2016), DOI: 10.1117/12.2234308

\bibitem{kpop}
K. U-Yen et al., {\it IEEE Transactions on Microwave Theory and Techniques} \textbf{56}, 172-177, (2008), DOI: 10.1109/TMTT.2007.912213

\bibitem{karwan14}
K. Rostem et al., {\it J. Appl. Phys.} \textbf{115}, 124508, (2014), DOI: 10.1063/1.4869737

\bibitem{wollack}
E. J. Wollack et al., {\it IEEE MTT-S International Microwave Symposium} \textbf{1}, 177-180, (2010), DOI: 10.1109/MWSYM.2010.5517063

\bibitem{crowe}
E. J. Crowe et al., {\it IEEE Transactions on Applied Superconductivity} \textbf{23}, 2500505, (2013), DOI:  10.1109/TASC.2012.2237211

\bibitem{aamir}
A. M. Ali et al., {\it Proc. SPIE} \textbf{10708}, 107082P, (2018), DOI: 10.1117/12.2312817

\bibitem{nist}
K. Irwin et al., {\it J. Low Temp. Phys. } \textbf{167}, 588-594, (2012), DOI: 10.1007/s10909-012-0586-7

\bibitem{mce}
E. S. Battistelli et al., {\it J. Low Temp. Phys.} \textbf{151}, 908-914, (2008), DOI: 10.1007/s10909-008-9772-z

\bibitem{atmosphere}
J. Pardo et al., {\it IEEE Transactions on Antennas and Propagation} \textbf{49}, 12, (2001), DOI: 10.1109/8.982447

\bibitem{fts}
Z. Pan et al., {\it arXiv:} \textbf{1905.07399}, (2019)

\bibitem{katie}
K. Harrington et al., {\it Proc. SPIE} \textbf{10708}, 107082M, (2018), DOI: 10.1117/12.2313614

\bibitem{tom}
T. Essinger-Hileman et al., {\it Proc. SPIE} \textbf{9153}, 91531I, (2014), DOI: 10.1117/12.2056701

\end{thebibliography}
\end{document}